\documentclass[letterpaper, conference]{IEEEtran}

\usepackage[utf8]{inputenc}

\IEEEoverridecommandlockouts

\usepackage{amssymb}
\usepackage{amsmath}
\usepackage{graphicx}
\usepackage{subcaption}
\usepackage{lscape}

\newif\ifcomment
\commenttrue
\commentfalse
\ifcomment
\newcommand{\comments}[1]{{\leavevmode\color{blue}#1}}
\else
\newcommand{\comments}[1]{}
\fi

\newcommand{\vsfig}[1]{\vspace{-0.0cm}}

\newcommand{\fref}[1]{Figure~\ref{#1}}
\newcommand{\tref}[1]{Table~\ref{#1}}
\newcommand{\eref}[1]{(\ref{#1})}

\DeclareMathOperator*{\argmax}{arg\,max}

\begin{document}


\title{A Markov Decision Process Model to Guide Treatment of Abdominal Aortic
Aneurysms\hspace{1mm}$^{\scriptscriptstyle\natural}$}

\author{\IEEEauthorblockN{Robert Mattila\IEEEauthorrefmark{1}$^{\wr}$,
Antti Siika\IEEEauthorrefmark{2}$^{\wr}$,
Joy Roy\IEEEauthorrefmark{2} and
Bo Wahlberg\IEEEauthorrefmark{1}
}

\thanks{This work was partially supported by the Swedish Research
Council, the Linnaeus Center ACCESS at KTH, the Swedish Heart-Lung Foundation and the Health, Technology
and Medicine grant from the Stockholm County Council. 
$\quad$\IEEEauthorrefmark{1}Robert Mattila and Bo Wahlberg are with the Department of Automatic Control,
School of Electrical Engineering, KTH Royal Institute of Technology,
Stockholm, Sweden. e-mails: \texttt{rmattila@kth.se, bo.wahlberg@ee.kth.se}
$\quad$\IEEEauthorrefmark{2}Antti Siika and Joy Roy are with the Department of Molecular Medicine and Surgery and the Department of Vascular Surgery,
Karolinska Institute and University Hospital, Stockholm, Sweden. e-mails:
\texttt{antti.siika@ki.se}, \texttt{joy.roy@ki.se}
$\quad$$^{\wr}$The authors wish it to be known that, in their opinion,
the first two authors should be regarded as joint First Authors.
$\quad$$^{\natural}$This work has been published, see \cite{mattila_markov_2016}.}}

\maketitle
\thispagestyle{empty}

\begin{abstract}

  An \emph{abdominal aortic aneurysm} (AAA) is an enlargement of the abdominal
  aorta which, if left untreated, can progressively widen and may rupture with
  fatal consequences.  In this paper, we determine an optimal treatment policy
  using Markov decision process modeling. The policy is optimal with respect to
  the number of \emph{quality adjusted life-years} (QALYs) that are expected to
  be accumulated during the remaining life of a patient. The new policy takes
  into account factors that are ignored by the current clinical policy (e.g. the
  life-expectancy and the age-dependent surgical mortality).  The resulting
  optimal policy is structurally different from the current policy. In
  particular, the policy suggests that young patients with small aneurysms
  should undergo surgery. The robustness of the policy structure is
  demonstrated using simulations. A gain in the number of expected QALYs is
  shown, which indicates a possibility of improved care for patients with AAAs.

\end{abstract}


\section{Introduction}

For any kind of surgical intervention, the risks involved in the operation must
be weighed against the potential benefits. In this paper, we focus on the
question of when an \emph{abdominal aortic aneurysm} (AAA) should be treated.
AAAs are enlargements of the aorta that in general are asymptomatic. They,
however, pose a threat of fatal rupture. Surgery aims to prevent this
rupture, and thereby maximize the remaining life expectancy of the patient. The
procedure is major surgery and can itself lead to death.

Each year during the lifetime of a patient with an AAA, the surgeon is faced
with the question ``should surgery be performed?'' Since the diameter of an
aneurysm is closely related to its rupture risk, the current clinical
guidelines for treatment of AAAs state that surgery is recommended if the
maximal aortic diameter exceeds 55 mm \cite{Moll2011}. This treatment policy is not
patient-specific, in the sense that it does not take into account factors such
as the life-expectancy or the expected surgical mortality of the patient.

The question that the surgeon faces can be seen as a problem of
sequential decision making under uncertainty. Such problems have previously
been studied in the control theory and operations research communities
under the name \emph{Markov decisions processes} (MDPs), see, e.g.,
\cite{puterman_markov_1994} or \cite{bertsekas_dynamic_2000}.

In this paper, the course of events and decisions resulting in the (potential)
rupture or treatment of an AAA is modeled using an MDP.
The resulting policy can guide the choice of treatment on a per patient basis:
Given the diameter of the AAA and the age of the patient, an \emph{optimal}
choice of treatment can be read from a table that is presented in
Section~\ref{sec:results}. The table is the result of optimizing the expected
number of \emph{quality adjusted life-years} (QALYs) over the patient's
remaining lifetime, where age dependent surgical and background mortalities have
been taken into account. The robustness of the policy, with respect to the
uncertainty in the model parameters, is assessed through simulations.

The main contributions of this paper are two-fold:
\begin{itemize}
    \item Firstly, we demonstrate how the treatment of AAA can be modeled using
    the MDP framework.
    \item Secondly, and the main conclusion, is that the policy currently
    employed in clinical practice does not have an optimal structure.
\end{itemize}

The rest of this paper is organized as follows. Section~\ref{sec:background}
presents a brief overview of AAAs and the QALY measure. It then proceeds with
a general outline and the notation of MDPs. In Section~\ref{sec:model} it is
explained how the problem allows an MDP formulation, as well as a discussion of
the numerical parameters used in the model. Section~\ref{sec:results} presents
the computed optimal policy and asserts the robustness.
Section~\ref{sec:discussion} concludes the paper with a discussion of related
work and indications for further extensions.

\section{Background and Preliminaries}
\label{sec:background}

\subsection{Abdominal aortic aneurysms}

\begin{figure}[t!]
  \centering
  \includegraphics[width=0.46\columnwidth]{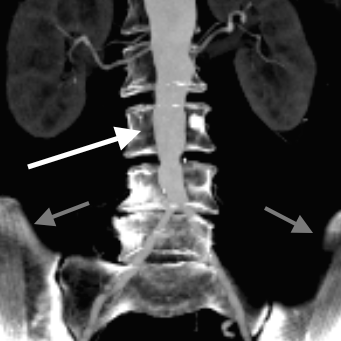}
  \hspace{0.02\columnwidth}
  \includegraphics[width=0.46\columnwidth]{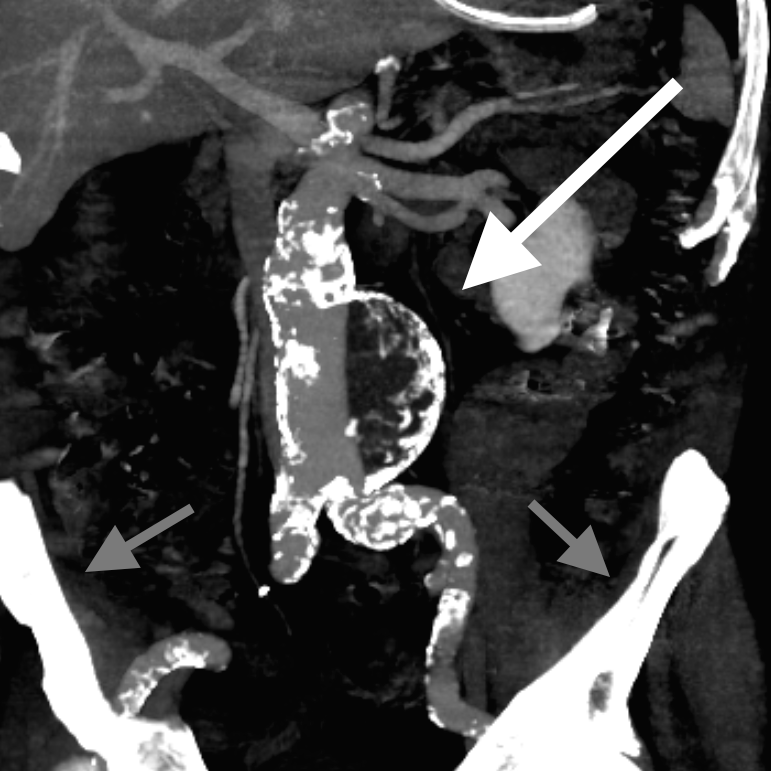}
  \caption{\emph{Computed tomography} (CT) angiograms depicting, with white
  arrows, a normal aorta (left) and an AAA (right). For reference, the gray
  arrows point at the hip bones in both figures. The difference in shading and
  visible organs is due to the method used to generate the figures.}
  \label{fig:aaa_picture}
  \vsfig{}
\end{figure}

An aneurysm is a balloon-like dilatation of an artery \cite{Moll2011}. They can
occur in all arteries, but are most common in the infra-renal aorta and are
there referred to as AAAs.  \fref{fig:aaa_picture} depicts \emph{computerized
tomography} (CT) angiograms of a normal non-aneurysmatic aorta and an AAA.

The disease is characterized by loss of structural integrity in the wall of the
abdominal aorta \cite{Sakalihasan2005}. This, in most cases, leads to a
progressive widening of the aneurysmal dilatation, that can potentially rupture
\cite{Sakalihasan2005}. The rupture of an AAA leads to massive hemorrhage and is
a medical emergency. It is fatal in up to 88\% of cases and 50\% of patients die
before reaching the hospital where they can undergo acute surgery
\cite{Bengtsson1993}. As many as 15 000 deaths each year in the USA are
attributable to AAAs \cite{Sakalihasan2005}.

It is possible to treat the AAA prior to its rupture (elective repair). Two
different techniques exist for the treatment:  \emph{open surgical repair} (OSR)
and \emph{endovascular aortic repair} (EVAR). In OSR the diseased part of the
aorta is replaced with a synthetic graft.  After the surgery, the patient does
not require any special follow-up, and is in essence cured from the disease
\cite{Sakalihasan2005}.  In the EVAR technique, a stent-graft is placed inside
the AAA through the arteries of the legs. This procedure requires that the
patient is regularly followed up post-operatively to monitor the development of
complications \cite{Sakalihasan2005}. There is no consensus on  which technique
is superior. Several studies have shown similar long-term mortality with
both techniques, see \cite{Sakalihasan2005} and \cite{Moll2011}. Instead, the choice
depends on the anatomy of the aneurysm, the health status of the patient, the
experience of the surgeon and patient preference \cite{Moll2011}.

The risk of rupture of an AAA is closely related to the aneurysm diameter.
Guidelines state that AAAs should be operated if the diameter of the aneurysm
exceeds 55 mm \cite{Moll2011}. This exact threshold of 55 mm was chosen based on
expert consensus (see \cite{Powell1993} and \cite{UKSAT1998}): All surgeons
would operate patients who had an aneurysm that exceeded 60 mm, but there was no
clear consensus on the policy for  smaller AAAs. This criterion has by
randomized controlled trials been shown to be superior to a strategy where also
small AAAs are operated \cite{Moll2011}.

The current policy, however, is
not perfect.  It was recently shown that a significant proportion of AAA rupture
before reaching the operative diameter threshold of 55 mm \cite{Laine2016}. It
is also known that some AAAs remain intact even after reaching a considerable
diameter (\textgreater 80mm), see \cite{Laine2016} and \cite{Heikkinen}.


There are patient-specific exceptions to the 55 mm policy. It is generally
recommended that women, who have larger risk of rupture, are operated when they
reach 52 mm instead of 55 mm \cite{Moll2011}. Also, most surgeons refrain from
intervention if the patient suffers from serious co-morbidities (e.g. terminal
cancer or heart disease) as this may increase the operative risk, patient
suffering or not be of any conceivable benefit to the patient. But, no rigorous
method exists for supporting the construction of these exceptions.

For an extensive review on the subject of AAAs, see \cite{Sakalihasan2005}.

\subsection{Quality adjusted life-years}

For a policy to be optimal it must specified with respect to what measure.
A \emph{quality adjusted life-year} (QALY) is a measure that describes the
quality of life. Living one year at perfect health is equivalent to one QALY,
and conversely one year dead is equivalent to zero QALYs. QALYs are used to
provide more nuanced descriptions of health states compared to simply counting
life-years, and are therefore commonly used in health-economic and outcomes
research \cite{Brazier1999}.

\subsection{Markov decision processes}

A widely used framework to model autonomous discrete-time stochastic systems is
the Markov chain model (see, e.g., \cite{cinlar_introduction_1975}).  In this
model, the system is assumed to transition randomly between a countable
number of possible states.  An MDP is an extension of this model in which the evolution
of the system can be influenced by the choice of an action at each time
instant. Solving an MDP means finding an optimal action to take, with respect to
some reward function, at each possible state of the system and at each point in
time. Such a mapping, not necessarily optimal, is called a \emph{policy}.

It has previously been demonstrated that the time-evolution of an AAA
(including growth and rupture risks) can be modeled using a Markov chain, see
\cite{thompson:brown} and \cite{soogard:laustsen}. A natural extension, and the
topic of the present work, is to formulate the decision making problem for AAAs
-- whether to take an action (i.e., perform surgery) or not -- as an MDP.

Formally, we let the state of the Markov chain at time $k$ be $x_k$ which
resides in the state-space $\mathcal{X}$ with $X$ elements. The transition
probabilities are given by
\[
    p_{ij}(u, k) = \Pr[x_{k+1} = j | x_k = i, u_k = u],
\]
where $i, j \in \mathcal{X}$ are states, $u \in \mathcal{U}$ is an action and
$\mathcal{U}$ is a (finite) set of available actions. The interpretation is
that the transition probability $p_{ij}(u, k)$ denotes the probability that the
system will be in state $j$ at the next time instant, given that it currently
(time $k$) is in state $i$ and that the action $u$ has been chosen. The set
$\mathcal{U}$ can in general be time and/or state dependent, but it will be
sufficient for our purposes to only consider a constant set.

The aim of the MDP is to find actions that maximize the expectation of some
accumulative reward.  We let the reward acquired at each time instant be
a mapping $r : \mathcal{X} \times \mathcal{U} \times \{M, M+1, \dots, N\}
\rightarrow \mathbb{R}$, where $N - M$ is the length of the time horizon over
which we consider the sequential decision problem. We interpret $r(i, u, k)$ as
the reward acquired by applying action $u$ at time $k$ when the system is in
state $i$. We assume that the terminal reward $r(i, \cdot, N)$ is independent
of the action chosen.

A policy is a sequence of functions $\pi = \{ \mu_M, \dots, \mu_{N-1} \}$,
where each $\mu_k$ maps a state to an action in the action set $\mathcal{U}$.
The expected (total) reward, if the system starts in state $x$, over a finite
time-horizon $N - M$ when applying the policy $\pi$ is
\[
    J_\pi ( x ) = \mathbb{E} \Big[ \; r(x_N, \cdot, N) + \sum_{k = M}^{N-1}
    r(x_k, \mu_k(x_k), k) \; \Big | x_M = x \Big].
\]
The solution of an MDP is a policy $\pi^*$ such that
\begin{equation}
    \pi^* = \argmax_\pi J_\pi (x).
    \label{eq:optimal_policy}
\end{equation}
In words, the optimal policy $\pi^*$ tells us what action we should take at
each time given the current state of the system, as to maximize the expected
accumulated reward during the remainder of the time-horizon over which we are
considering the problem. Note that any policy, in our setting, can be
illustrated in a look-up table since the state-space and time-horizon are
finite.

\begin{figure*}[t!]
\centering
\begin{subfigure}[t]{2.0\columnwidth}
    \centering
    \includegraphics[width=1\linewidth]{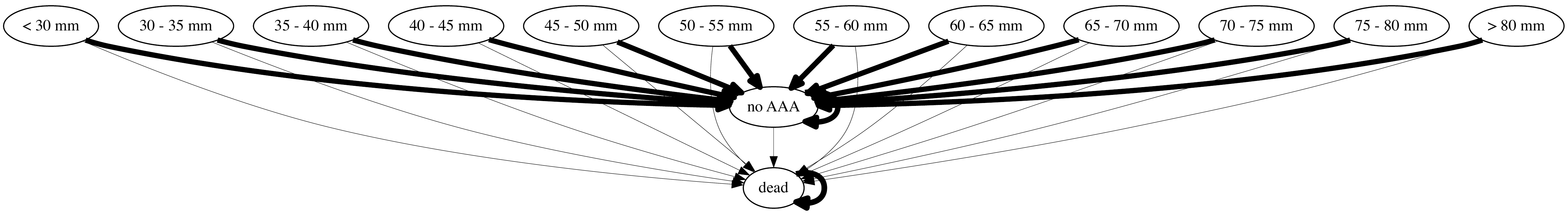}
    \caption{Action: perform surgery.}
    \label{fig:op}
\end{subfigure}
\begin{subfigure}[t]{1.3\columnwidth}
    \centering
    \includegraphics[width=1\linewidth, height=10.5cm]{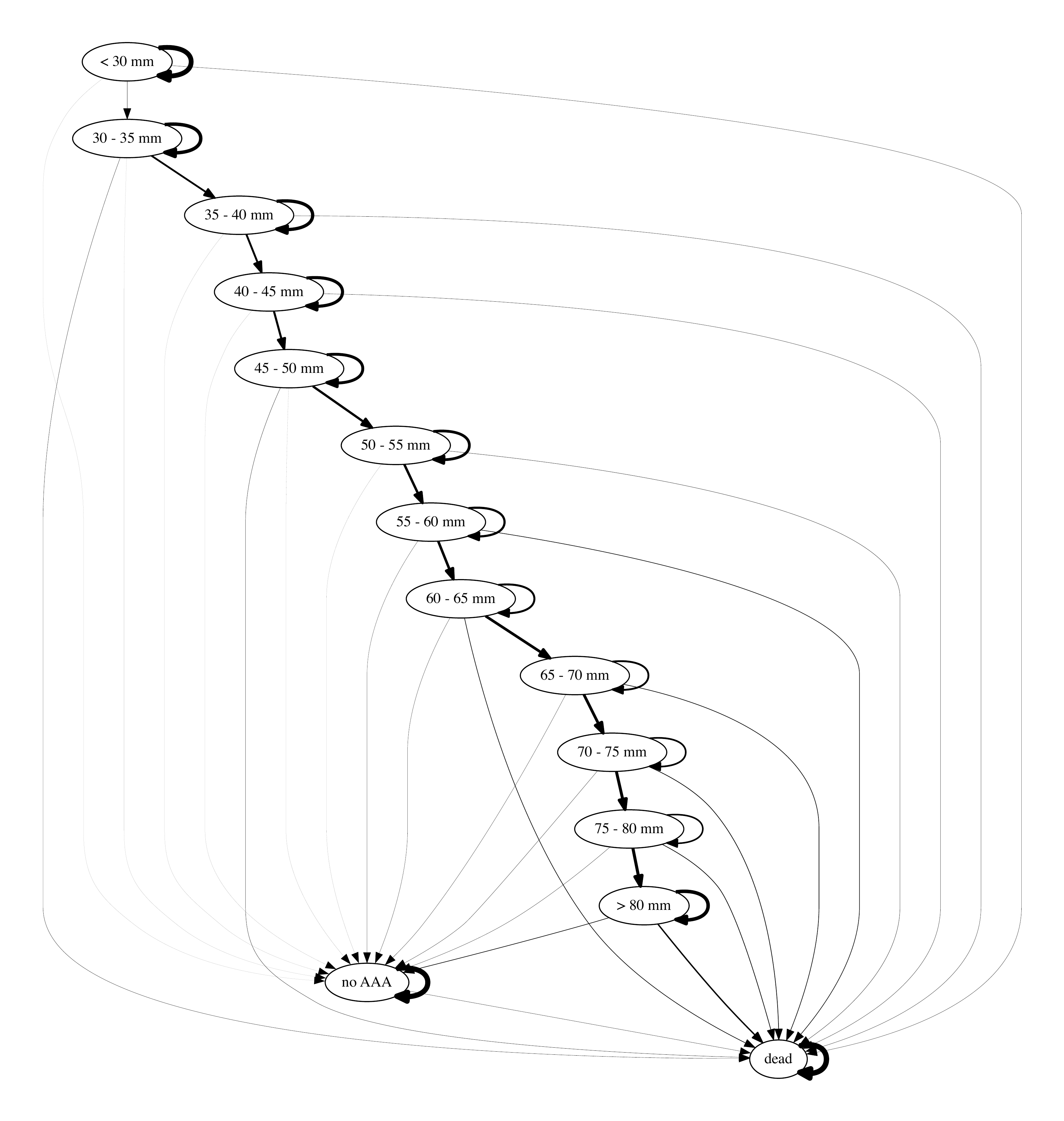}
    \caption{Action: continue surveillance.}
    \label{fig:no_op}
\end{subfigure}
\caption{Structure of the MDP illustrating the two Markov chains resulting from
    the two available actions; perform surgery and continue surveillance (for
    one year). The (age-dependent)
probability of a transition is proportional to the thickness of the associated
arrow. }
\label{fig:mdp_structure}
  \vsfig{}
\end{figure*}

A complete treatment of MDPs, along with methods for solving problem
\eref{eq:optimal_policy}, can be found in any of the standard textbooks; e.g.,
\cite{puterman_markov_1994}, \cite{bertsekas_dynamic_2000}.

\section{Model}
\label{sec:model}

\begin{figure*}[]
\centering
\begin{subfigure}[t]{0.45\textwidth}
    \centering
    \includegraphics[width=1.0\columnwidth]{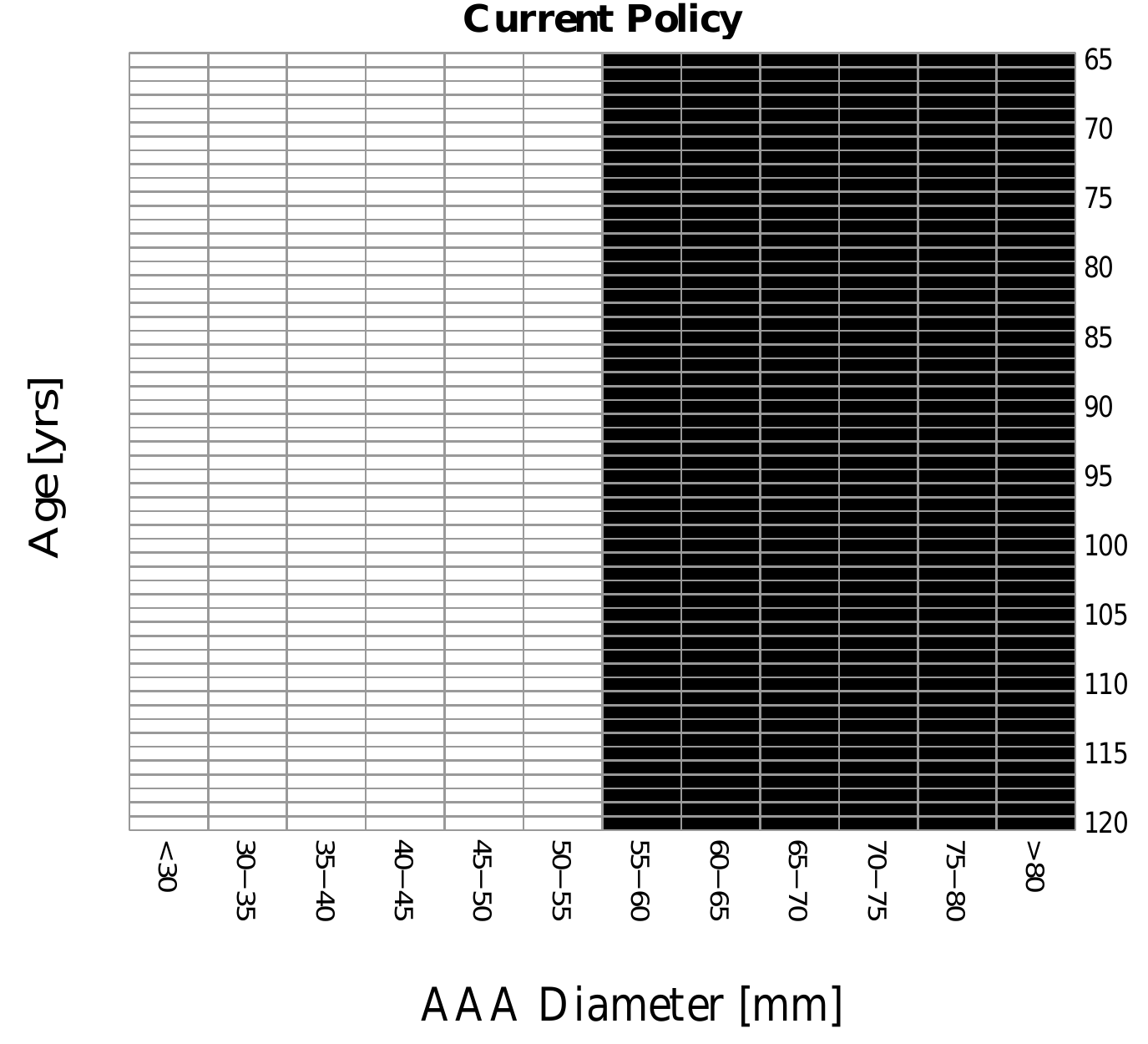}
    \caption{Current clinical policy, $\pi_{55}$.}
    \label{fig:policy_c}
\end{subfigure}
\begin{subfigure}[t]{0.05\textwidth}
\hspace{0.1cm}
\end{subfigure}
\begin{subfigure}[t]{0.45\textwidth}
    \centering
    \includegraphics[width=1.0\columnwidth]{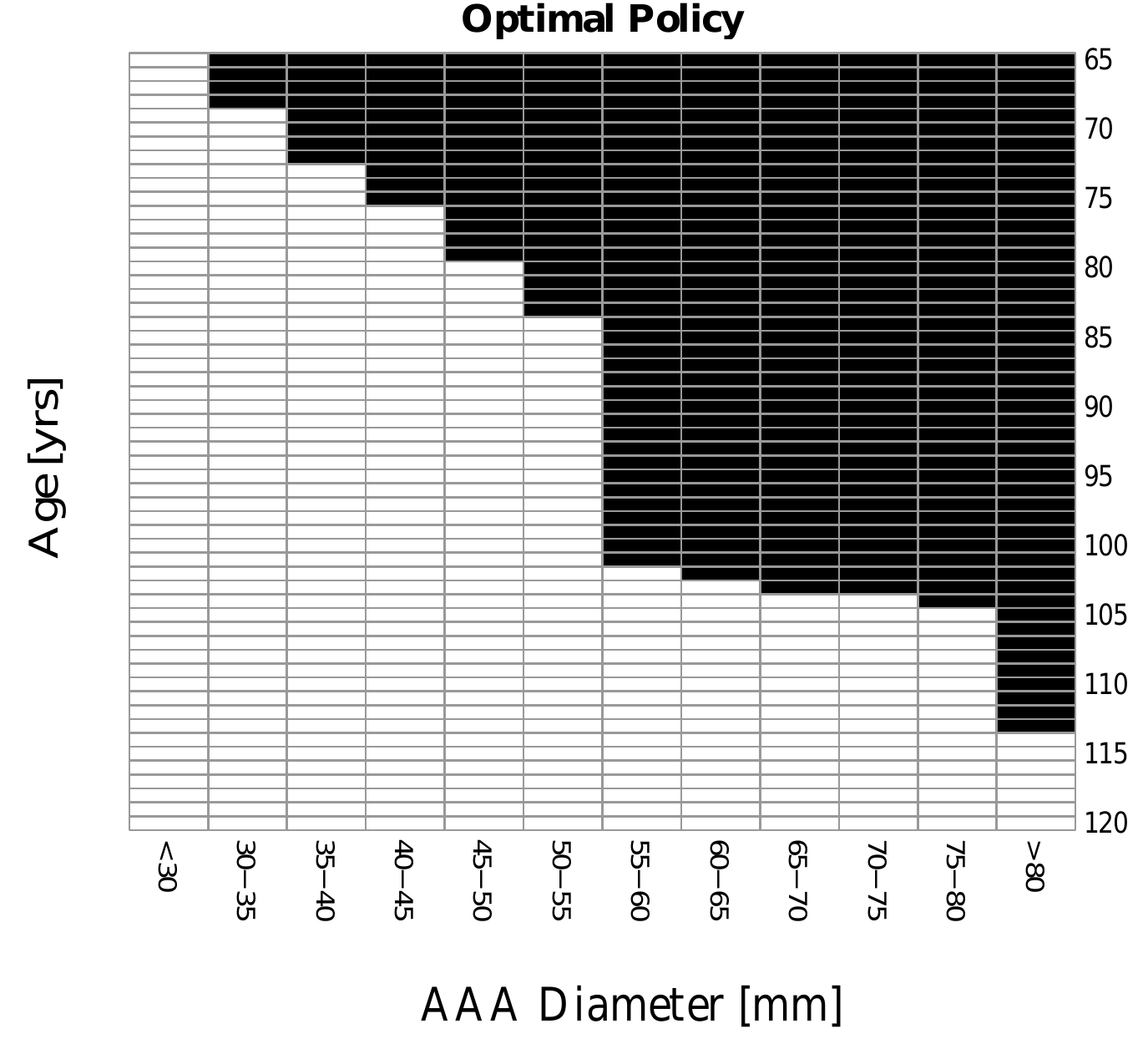}
    \caption{Optimal policy, $\pi^*$.}
    \label{fig:policy}
\end{subfigure}

\caption{The figures illustrate the current clinical treatment policy and the
    calculated optimal policy. The horizontal axes show size of the AAA and the
    vertical axes show age of the patient. A black cell indicates that surgery
    should be performed and a white cell indicates that no action should be
    taken (i.e., continue surveillance for one year). }
\label{fig:policies}

\vsfig{}

\end{figure*}

To formulate the AAA treatment problem in the above framework, we need to define the
state-space, the reward function, the action set and the transition
probabilities for an AAA. The objective is to determine an optimal
policy $\pi^*$ that will maximize the expected number of QALYs to be
accumulated over the patient's remaining life. We make the assumption that
there is a maximal age of $N = 120$ years and that the interesting interval for
optimization is from an age of $M = 65$ years.

Due to available data, we let the time-step of our MDP be one year. The reward
function $r(i,u,k)$ is defined to be the QALY equivalent of the age $k$ if the
patient is alive, and zero otherwise.  In other words, the reward gained in one
year is one year compensated for the age of the patient. It should be noted
that it is not at all obvious how this compensation should be calculated,
however, that is the subject of other works, see, e.g., \cite{Brazier1999} and
\cite{soogard:laustsen}. We let the quantized diameter of the AAA be the state
of the system. The interval of quantization was chosen as 5 mm due to the
available data (see the discussion on rupture and growth probabilities below). To
make the MDP terminate in case of death by rupture, surgery or natural causes,
we introduce an auxiliary terminal state (label: dead) and also a state for the
system after successful surgery (label: no AAA). The state-space is therefore
the set
\begin{align}
    \mathcal{X}
    = \{ \text{dead},& \text{ no AAA}, <30 \text{ mm}, 30-35\text{ mm}, \notag
    \\
&\dots, 70-75\text{ mm}, 75-80\text{ mm}, >80\text{ mm}\}.
\end{align}

We allow for two actions in our MDP model: $\mathcal{U} = \{\text{continue
surveillance}, \text{ perform surgery}\}$. These two actions influence
the transition probabilities of the system as can be seen in
Figures~\ref{fig:op} and \ref{fig:no_op}.  Figure~\ref{fig:op} illustrates the
structure of the model if the choice of action is to perform surgery and
Figure~\ref{fig:no_op} shows the structure of the model if the choice of action
is to not perform surgery (i.e., continue surveillance for one year).

To be able to compare our results to the current clinical policy, i.e., perform
surgery if the diameter of the AAA is greater than 55 mm, it is illustrative to
show how the current policy can expressed in the MDP framework. The policy
$\pi_{55} = \{\mu_M^{55}, \dots, \mu_{N-1}^{55}\}$ has $\mu_i^{55} = \mu^{55}$
for all $i = M, M+1, \dots, N-1$ where
\[
    \mu^{55}(x) = \begin{cases}
        \substack{\text{\normalsize{continue}} \\ \text{\normalsize{surveillance}}} &\text{if } x \in
        \{\text{dead, no AAA, } \\
& \hspace{0.0cm} \text{$<$ 30 mm,}  \dots\text{, 50-55mm} \}, \\
            \substack{\hspace{0.3cm}\text{\normalsize{perform}} \\ \;\text{\normalsize{ surgery}}} &\text{ otherwise}. \\
               \end{cases}
\]
The policy $\pi_{55}$ is graphically illustrated in Figure~\ref{fig:policy_c}
where a black cell indicates that surgery should be performed and a white cell
indicates that no intervention should be made.

Numerical values for the transition probabilities between different states and
the reward function were obtained from the available literature on the topic,
as discussed in the following.  AAA is a disease that is most prevalent in
males, and therefore most literature, and our scope, is limited to male
patients. \cite{soogard:laustsen} synthesized evidence regarding aneurysm
growth, rupture risk and age-dependent modeling of QALYs for a Markov chain
model which evaluated the potential benefits from screening as compared to
non-screening for AAA. The same parameters were used in our model. However, in
\cite{soogard:laustsen}, no rupture probabilities for AAAs that are smaller
than 50 mm are reported. Hence, rupture probabilities for aneurysms with
a diameter below 50 mm were retrieved from a systematic meta-analysis that
complied rupture probabilities from 14 studies of small aneurysms
\cite{thompson:brown}.

Instead of the case of a ruptured AAA leading to certain death, we included in
our modeling the chance of reaching a hospital  and there undergoing successful
emergency surgery. We considered only the surgical intervention called
\emph{open surgical repair} (OSR) of AAAs. Parameters, i.e.,
age-dependent surgical mortalities for both emergency and elective settings, as
well as the probability of reaching a hospital, were retrieved from
\cite{TheSwedishNationalRegistryforVascularSurgery2012}. Data for age-dependent
background morality was collected from \cite{SocialSecurityAdministration2010}.
The above discussed parameters and their references are summarized in
\tref{tbl:summary}.

\begin{table}[b!]
\begin{center}
\resizebox{\columnwidth}{!}{
\begin{tabular}{| c | c || c | c |}
    \hline
    Parameter & Reference & Parameter & Reference \\
    \hline
    Rupture probabilities & \cite{thompson:brown}, \cite{soogard:laustsen} & Growth
    probabilities & \cite{soogard:laustsen}  \\
    QALY, $c$  & \cite{soogard:laustsen} & Surgical mortalities & \cite{TheSwedishNationalRegistryforVascularSurgery2012} \\
    Reaching hospital & \cite{TheSwedishNationalRegistryforVascularSurgery2012}
    & Background mortalities & \cite{SocialSecurityAdministration2010} \\
    \hline
\end{tabular}
}
\end{center}
\caption{Summary of the parameters used in the model and the corresponding
references.}
\label{tbl:summary}
\end{table}

\section{Results}
\label{sec:results}

Problem \eref{eq:optimal_policy} was solved for the model outlined in the
previous section using a Python implementation of dynamic programming.  Data
analysis and generation of plots were performed using the R programming
language. Generating the optimal policy takes approximately one second on a 3.2
GHz MacBook Pro.

The resulting optimal policy is illustrated in Figure~\ref{fig:policy}. It is
clearly structurally different from the current policy which is illustrated in
Figure~\ref{fig:policy_c}. In these figures, the horizontal axes show the size
of the aneurysm, and the vertical axes show the age of the patient. A black
cell indicates that surgery should be performed, and a white cell indicates
that no action should be taken (i.e., continue surveillance for one year). The
optimal policy shows that younger patients benefit from surgery at small
diameters, and that the threshold diameter should be increased as the patient
ages.

\begin{figure}[ht!]
\centering
\begin{subfigure}[t]{0.43\textwidth}
    \centering
    \includegraphics[width=1.0\columnwidth]{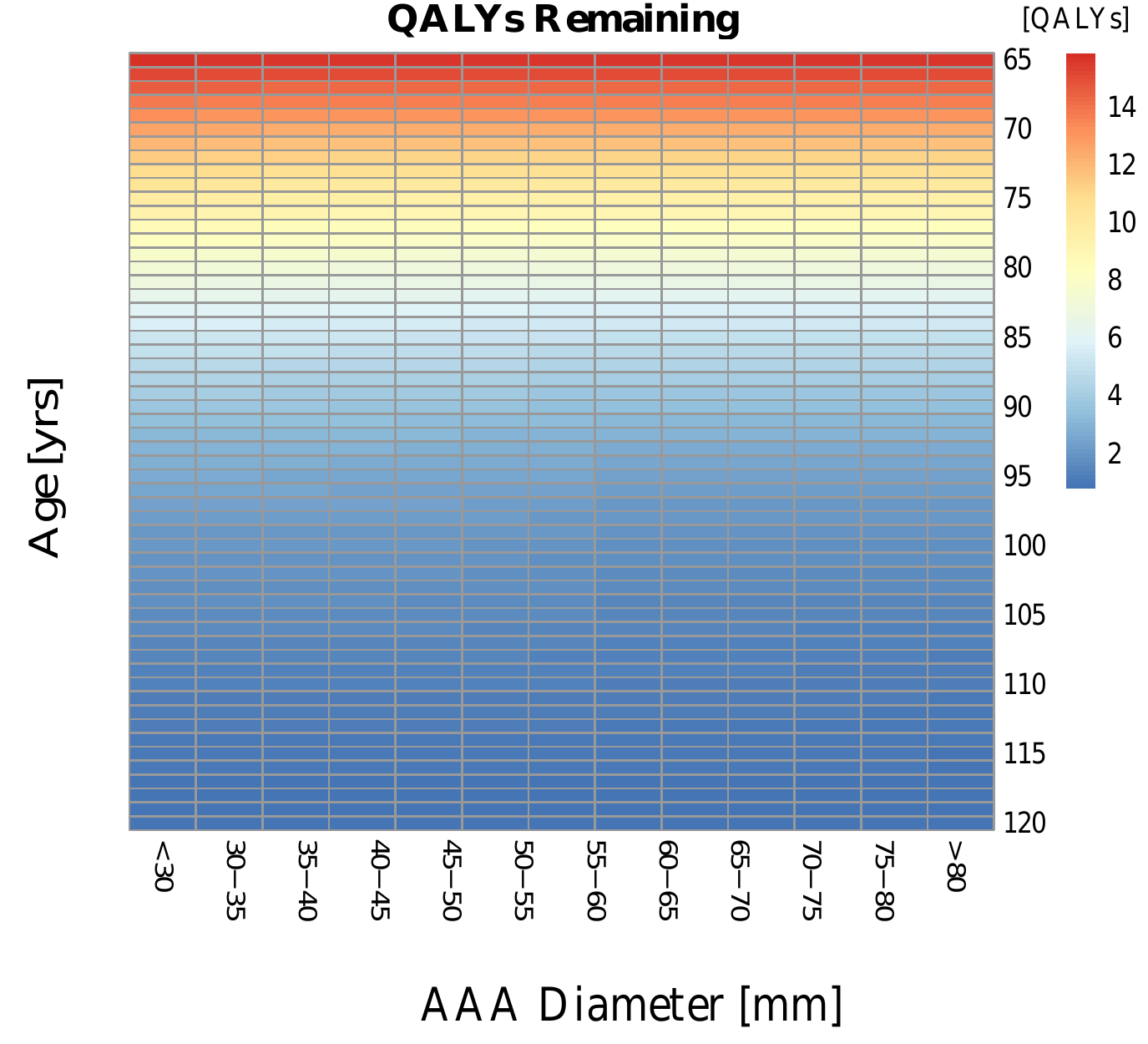}
    \caption{}
    \vspace{0.3cm}
    \label{fig:policy_qaly}
\end{subfigure}
\begin{subfigure}[t]{0.43\textwidth}
\centering
\includegraphics[width=1.0\columnwidth]{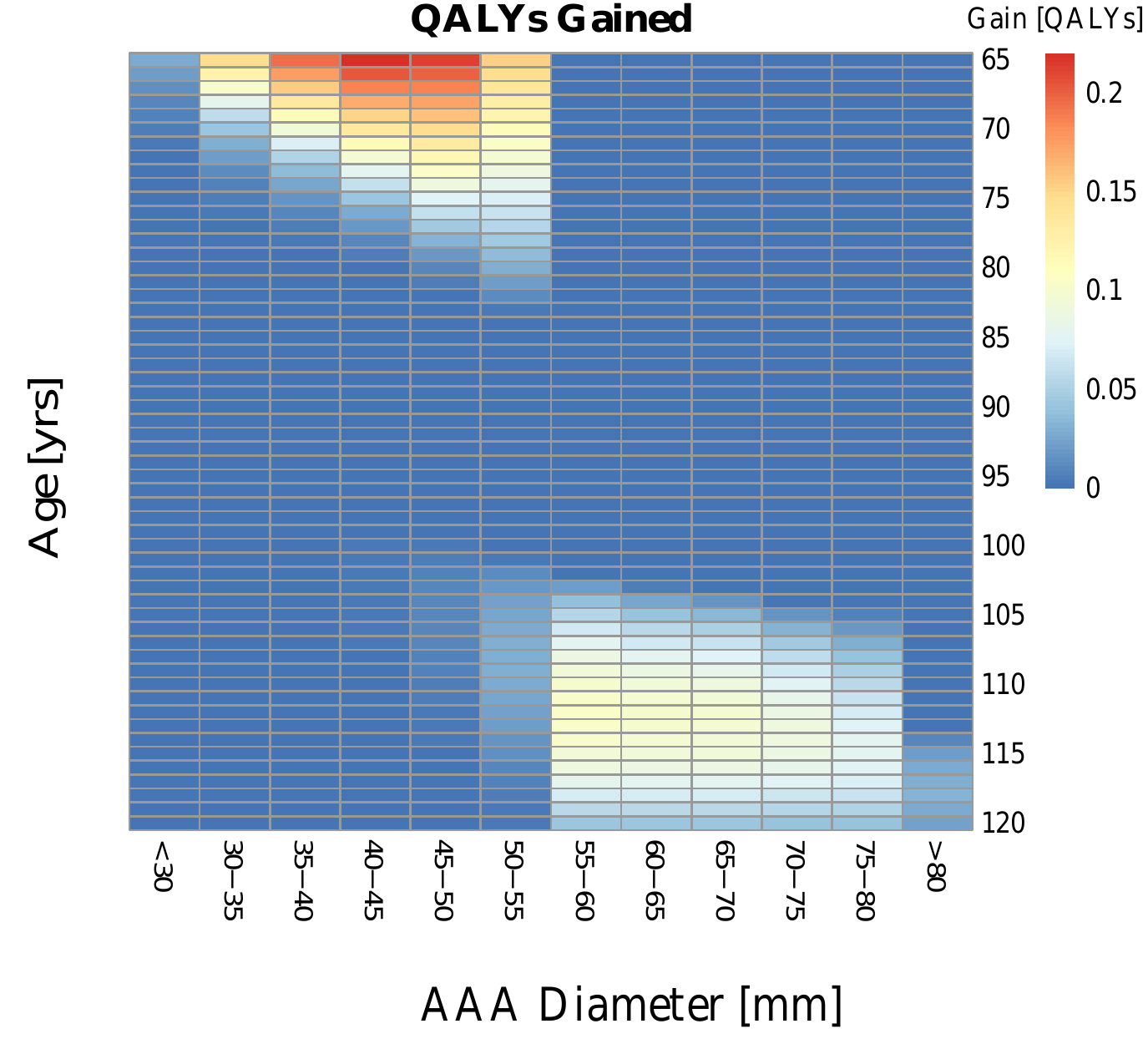}
\caption{}
\label{fig:gains}
\end{subfigure}

\caption{Color indicates the number of QALYs that either (a) can be expected
to be accumulated during the remaining life of the patient when the optimal
policy $\pi^*$ is employed, or (b) can be gained by
changing policy from the current clinical policy $\pi_{55}$ to the optimal policy $\pi^*$.
The horizontal axes show size of the AAA and
the vertical axes show age of the patient. }

\vsfig{}

\end{figure}

\fref{fig:policy_qaly} displays the total expected number of QALYs that are
expected to be accumulated  during the remainder of a patient's life  using the
optimal policy. \fref{fig:gains} shows the improvement in terms of the difference
in number of QALYs using the two policies in Figure~\ref{fig:policies}.

As indicated by Figure~\ref{fig:gains}, an improvement of the number of QALYs
can be expected. This improvement is most significant in patients aged 65 to 80
years with aneurysms sized 30 to 55 mm. There is also an improvement for older
patients aged 105 to 120 with aneurysms sized 55 to 80 mm and above. The relevance
for such old patients is, however, primarily hypothetical. These
improvements are expected since the optimal policy differ from the current one
for these patients. A small improvement in the number of QALYs can
also be expected in the region that corresponds to patients aged 105-115 with
aneurysms that are 50-55 mm in size. While the two policies here coincide, the
optimal policy assumes optimal action in the future (when the AAA grows), that
leads to an improvement.


There is a noticeable difference in the proposed policy with respect to at which
age operation is the optimal treatment choice between aneurysms that are smaller
and larger than 55 mm. This may be due to the fact that the only high quality
evidence from controlled clinical trials for aneurysm growth and rupture rates
relate to small aneurysms, whereas for large aneurysms only smaller case
series with selected patient-cohorts are available. For an extended discussion,
see Appendix~\ref{app:big_AAA}.

\begin{figure}[b!]
    \centering
    \includegraphics[width=1.0\columnwidth]{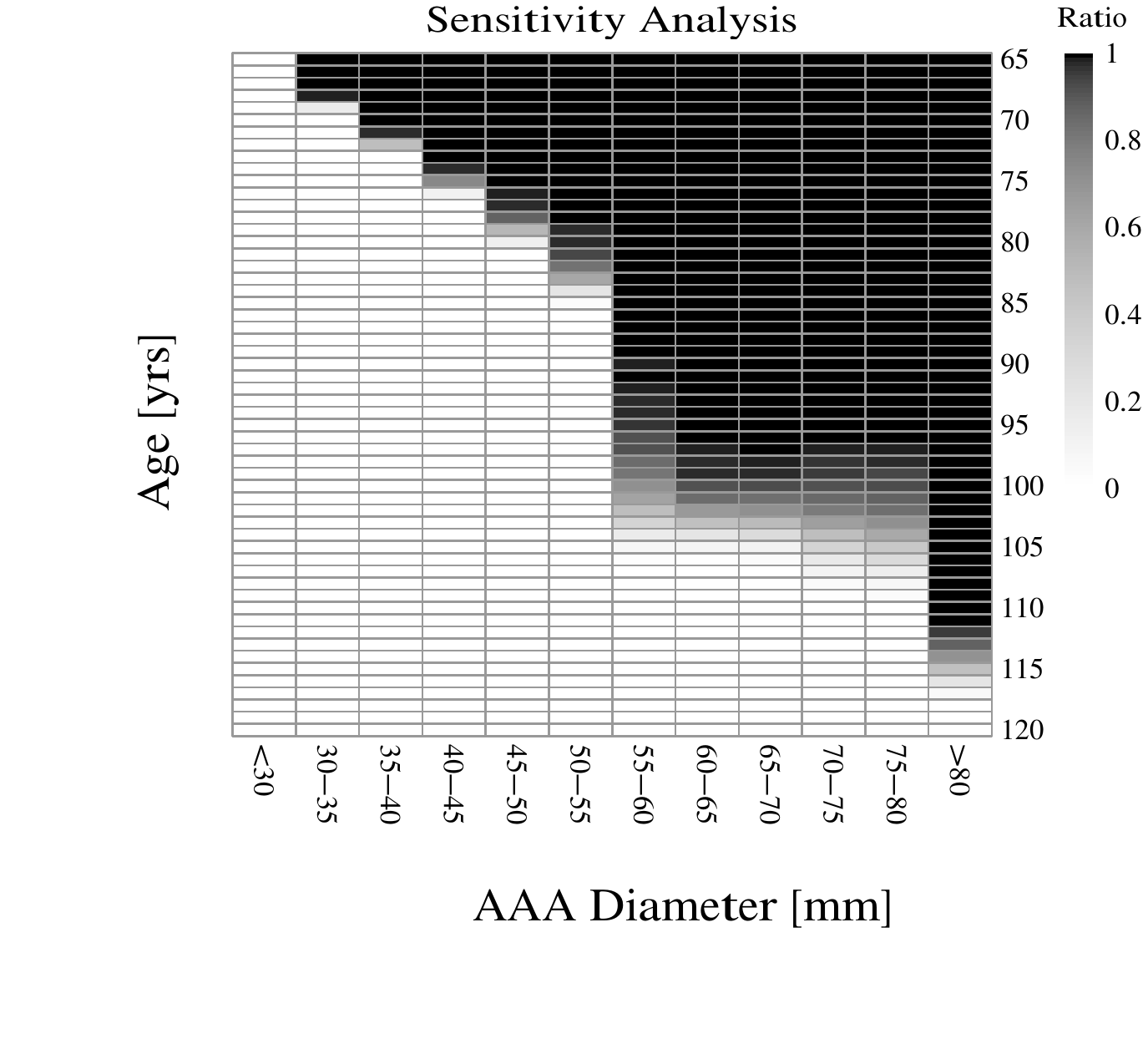}
    \caption{The ratios of policies that indicate that a certain action should be
    taken for 1000 randomly perturbed sets of model parameters.}
    \label{fig:policy_pert}
    \vsfig{}
\end{figure}

As always when using measured parameters in a model, it is of interest to
evaluate how robust the results are with respect to the uncertainty in the
parameters. To test the robustness of the calculated optimal policy, we
generated random perturbations from the published data uncertainties on the rupture
probabilities (from \cite{soogard:laustsen} and \cite{thompson:brown}). One
thousand sets of parameters were generated and optimal policies were calculated
for each set. The ratio of policies that indicated that a certain action should
be taken at a certain age and aneurysm size is shown in \fref{fig:policy_pert}.
It can be observed that the structural shape of the optimal policy is stable
with respect to the perturbations of the parameters.

\section{Discussion}
\label{sec:discussion}

We have shown that a policy that optimally takes into account the expected
remaining (quality adjusted) life-years of a patient and the age-related
surgical mortality  differs markedly from the policy which is currently used
in clinical practice. The structure of the suggested policy is intuitively
appealing: For a patient with less expected remaining life and a high operative
risk, a higher AAA rupture probability (size) is demanded in order to take
action.

The exact thresholds in the generated policy may not be clinically applicable
yet. They are subject to change as better estimates of the model parameters
become available from clinical studies. However, as shown, the structure of the
policy is robust to perturbations in the current model parameters which
indicates that even if the exact threshold values are not determined yet,
the structure of the currently used clinical policy is sub-optimal and
demands further investigation.

As previously mentioned in Section~\ref{sec:background}, the current policy has
been studied in randomized controlled trials and has been shown to be
associated with a lower mortality than a policy where small (40-55 mm)
aneurysms are also operated \cite{Moll2011}. These trials, however, did not
have sufficient power to perform sub-group analyses with regard to age or
gender \cite{Moll2011}, and could thus not demonstrate the results suggested
here, i.e., that smaller aneurysms should be operated in young patients.

There are many extensions that could be made to the work presented in this
paper. We considered only age-dependent surgical mortality that is related to
open surgical repair. It would be interesting to include, by extending the
action set $\mathcal{U}$, other types of surgical interventions.  A time or
state dependent $\mathcal{U}$ could limit some of these surgical methods to
certain ages or certain AAA diameters. Post-operative complications can also be
introduced in the model by the addition of new states.

It should be noted that, in this paper, age can partially be viewed as
a surrogate marker for co-morbidity.  It is possible to identify patients with
increased operative risk (due to co-morbidities) using, for example, the score
proposed in \cite{Ambler2015}, or patients with an increased rupture risk
(e.g., women or patients with a family history of ruptured AAAs), for whom
individual risk parameters can be used to generate patient-specific policies.
It may also be relevant to improve patient-centered decision making by allowing
patients themselves to estimate QALYs.

Also, as more specific markers for AAA rupture and growth (e.g., biomechanical
rupture risk markers from finite element analysis \cite{Martufi2013}) become
clinically available, policies will have to be re-established to include such
data as basis for the decision. Such new policies, we believe, should be
synthesized with the help of mathematical methods for decision-making. The MDP
framework admits new markers to be incorporated by either an extension of the
state-space or by extending the framework to \emph{partially observed MDPs} (POMDPs)
\cite{vikram2016}.

\subsection{Related work}

In the context of healthcare, the MDP framework
has been used to provide guidelines for decisions in such
disperse settings as ambulance scheduling \cite{maxwell_approximate_2010},
planning the treatment of ischemic heart disease
\cite{hauskrecht_planning_2000} and kidney transplantations
\cite{ahn_involving_1996}. For a more extensive overview of healthcare related
applications of MDPs, see for example  \cite{schaefer_modeling_2005} and
references therein.

The question of improving the treatment policy for AAAs using mathematical
modeling has been considered before. However, the MDP framework has, to the
knowledge of the authors, not been used before. The structure of our policy
is similar to the one recently obtained in \cite{grant} using a Markov model
where different treatment options were evaluated against each other by
simulating their outcome over a large cohort of (virtual) patients. However, in
contrast to enforcing a surgical intervention at different AAA diameters and
comparing the simulation outcomes, the MDP framework \emph{a)} uses analytical
expression for expectations, which bypasses the need for approximations using
simulations and makes the computation of the solution
more efficient. It also \emph{b)} generates a policy that takes into account that the
surgeon will act optimally in the future due to the \emph{principle of
optimality} \cite{bellman:1957}. This means that we include the possibility of
\emph{not} performing surgery now, since we know that it will be performed at a
later stage, when generating the policy. A more elaborate discussion regarding
this is available in \cite{magni_deciding_2000}. Moreover, we believe our
framework allows the extensions discussed in the previous section to be made
more easily.

\section{Conclusions}
\label{sec:conclusion}

In this paper we have demonstrated how methods from operations research on
sequential decision making can be employed for weighing risk against potential
benefit in the case of AAA treatment. We have demonstrated that
a patient-specific policy outperforms the currently used policy. Our results
indicate that the optimal treatment policy might be of a more complex form
- age and size dependent - than the one that is employed today. In particular,
smaller aneurysms should be operated in younger patients. These results warrant
further investigations into a policy that is age dependent.

\comments{
\section{Appendix A}

The method developed in this paper does not only apply to AAAs. It has been nicely novelized  in the
context of cerebral aneurysms in Henry Marsh's book \emph{Do No Harm},
as quoted below:
\begin{quote}
    ``But should I?''

    ``I don't know.''

    She was right. I didn't know either. If we did nothing the patient might
    eventually suffer a haemorrhage which would probably cause a catastrophic
    stroke or kill her. But then she might die years away from something else
    without the aneurysm ever having burst. She was perfectly well at the
    moment, the headaches for which she had had the scan were irrelevant and
    had got better. The aneurysm had been discovered by chance. If I operated
    I could cause a stroke and wreck her – the risk of that would probably be
    about four or five per cent. So the acute risk of operating was roughly
    similar to the life­time risk of doing nothing. Yet if we did nothing she
    would have to live with the knowledge that the aneurysm was sitting there
    in her brain and might kill her any moment.

    (...)

    ``What would you do if it was you?''

    I hesitated, but the fact of the matter was that by the age of sixty­one
    I was well past my best­by date and I knew that I had already lived most of
    my life. Besides, the difference in our ages meant that I had fewer years
    of life ahead of me so the life­time risk of the aneurysm rupturing, if it
    was not operated on, would be much lower for me and the relative risk of
    the operation correspondingly higher.
\end{quote}
}


\bibliographystyle{ieeetr}
\bibliography{references}

\begin{thebibliography}{10}

\bibitem{mattila_markov_2016}
R.~Mattila, A.~Siika, J.~Roy, and B.~Wahlberg, ``A {Markov} decision process
  model to guide treatment of abdominal aortic aneurysms,'' in {\em
  {Proceedings} of the {IEEE} {Conference} on {Control} {Applications}
  (CCA'16)}, pp.~436--441, sep 2016.

\bibitem{Moll2011}
F.~L. Moll, J.~T. Powell, G.~Fraedrich, F.~Verzini, S.~Haulon, M.~Waltham,
  J.~A. van Herwaarden, P.~J.~E. Holt, J.~W. van Keulen, B.~Rantner, F.~J.~V.
  Schl\"{o}sser, F.~Setacci, and J.-B. Ricco, ``{Management of abdominal aortic
  aneurysms clinical practice guidelines of the European society for vascular
  surgery.},'' {\em European Journal of Vascular and Endovascular Surgery},
  vol.~41 Suppl 1, pp.~S1--S58, Jan. 2011.

\bibitem{puterman_markov_1994}
M.~L. Puterman, {\em Markov Decision Processes: Discrete Stochastic Dynamic
  Programming}.
\newblock New York, NY, USA: John Wiley \& Sons, Inc., 1994.

\bibitem{bertsekas_dynamic_2000}
D.~P. Bertsekas, {\em Dynamic Programming and Optimal Control}.
\newblock Belmont, Mass.: Athena Scientific, 2nd~ed., 2000.

\bibitem{Sakalihasan2005}
N.~Sakalihasan, R.~Limet, and O.~D. Defawe, ``Abdominal aortic aneurysm,'' {\em
  The Lancet}, vol.~365, pp.~1577--89, jan 2005.

\bibitem{Bengtsson1993}
H.~Bengtsson and D.~Bergqvist, ``{Ruptured abdominal aortic aneurysm: a
  population-based study.},'' {\em Journal of Vascular Surgery}, vol.~18,
  pp.~74--80, jul 1993.

\bibitem{Powell1993}
J.~Powell, R.~Greenhalgh, C.~Ruckley, and F.~Fowkes, ``{Prologue to a surgical
  trial},'' {\em The Lancet}, vol.~342, pp.~1473--1474, dec 1993.

\bibitem{UKSAT1998}
``{Mortality results for randomised controlled trial of early elective surgery
  or ultrasonographic surveillance for small abdominal aortic aneurysms},''
  {\em The Lancet}, vol.~352, pp.~1649--1655, nov 1998.

\bibitem{Laine2016}
M.~T. Laine, T.~V{\"{a}}nttinen, I.~Kantonen, K.~Halmesm{\"{a}}ki, E.~M.
  Weselius, S.~Laukontaus, J.~Salenius, P.~S. Aho, and M.~Venermo, ``Rupture of
  abdominal aortic aneurysms in patients under screening age and elective
  repair threshold,'' {\em European Journal of Vascular and Endovascular
  Surgery}, vol.~51, pp.~511--516, feb 2016.

\bibitem{Heikkinen}
M.~Heikkinen, J.~Salenius, and O.~Auvinen, ``{Ruptured abdominal aortic
  aneurysm in a well-defined geographic area},'' {\em Journal of Vascular
  Surgery}, vol.~36, pp.~291--296, aug 2002.

\bibitem{Brazier1999}
J.~Brazier, M.~Deverill, C.~Green, R.~Harper, and A.~Booth, ``{A review of the
  use of health status measures in economic evaluation.},'' {\em Health
  Technology Assessment}, vol.~3, pp.~i--iv, 1--164, jan 1999.

\bibitem{cinlar_introduction_1975}
E.~Çinlar, {\em Introduction to Stochastic Processes}.
\newblock Englewood Cliffs, NJ: Prentice-Hall, 1975.

\bibitem{thompson:brown}
S.~G. Thompson, L.~C. Brown, M.~J. Sweeting, M.~J. Bown, L.~G. Kim, M.~J.
  Glover, M.~J. Buxton, and J.~T. Powell, ``Systematic review and meta-analysis
  of the growth and rupture rates of small abdominal aortic aneurysms:
  implications for surveillance intervals and their cost-effectiveness,'' {\em
  Health Technology Assessment}, vol.~17, no.~41, pp.~1--118, 2013.

\bibitem{soogard:laustsen}
R.~S{\o}gaard, J.~Laustsen, and J.~S. Lindholt, ``Cost effectiveness of
  abdominal aortic aneurysm screening and rescreening in men in a modern
  context: evaluation of a hypothetical cohort using a decision analytical
  model,'' {\em BMJ}, vol.~345, 2012.

\bibitem{TheSwedishNationalRegistryforVascularSurgery2012}
{Swedish Agency for Health Technology Assessment and Assessment of Social
  Services}, ``Screening för bukaortaaneurysm – rekommendation och
  bedömningsunderlag – remissversion,'' tech. rep., nov 2015.

\bibitem{SocialSecurityAdministration2010}
{Social Security Administration}, ``Annual statistical supplement to the social
  security bulletin, 2011 (actuarial life table).'' Retrieved from
  http://www.ssa.gov/oact/STATS/table4c6.html.
\newblock SSA Publication No. 13-11700.

\bibitem{Ambler2015}
G.~K. Ambler, M.~S. Gohel, D.~C. Mitchell, I.~M. Loftus, and J.~R. Boyle, ``The
  abdominal aortic aneurysm statistically corrected operative risk evaluation
  ({AAA SCORE}) for predicting mortality after open and endovascular
  interventions.,'' {\em Journal of Vascular Surgery}, vol.~61, pp.~35--43, jan
  2015.

\bibitem{Martufi2013}
G.~Martufi and T.~{Christian Gasser}, ``{Review: the role of biomechanical
  modeling in the rupture risk assessment for abdominal aortic aneurysms.},''
  {\em Journal of Biomechanical Engineering}, vol.~135, p.~021010, feb 2013.

\bibitem{vikram2016}
V.~Krishnamurthy, {\em Partially Observed Markov Decision Processes}.
\newblock Cambridge, United Kingdom: Cambridge University Press, 2016.

\bibitem{maxwell_approximate_2010}
M.~S. Maxwell, M.~Restrepo, S.~G. Henderson, and H.~Topaloglu, ``Approximate
  dynamic programming for ambulance redeployment,'' {\em INFORMS Journal on
  Computing}, vol.~22, pp.~266--281, May 2010.

\bibitem{hauskrecht_planning_2000}
M.~Hauskrecht and H.~Fraser, ``Planning treatment of ischemic heart disease
  with partially observable {Markov} decision processes,'' {\em Artificial
  Intelligence in Medicine}, vol.~18, no.~3, pp.~221--244, 2000.

\bibitem{ahn_involving_1996}
J.-H. Ahn and J.~C. Hornberger, ``Involving patients in the cadaveric kidney
  transplant allocation process: {A} decision-theoretic perspective,'' {\em
  Management Science}, vol.~42, no.~5, pp.~629--641, 1996.

\bibitem{schaefer_modeling_2005}
A.~J. Schaefer, M.~D. Bailey, S.~M. Shechter, and M.~S. Roberts, ``Modeling
  medical treatment using {Markov} decision processes,'' in {\em Operations
  Research and Health Care}, pp.~593--612, Springer, 2005.

\bibitem{grant}
S.~W. Grant, M.~Sperrin, E.~Carlson, N.~Chinai, D.~Ntais, M.~Hamilton, G.~Dunn,
  I.~Buchan, L.~Davies, and C.~N. McCollum, ``Calculating when elective
  abdominal aortic aneurysm repair improves survival for individual patients:
  development of the {Aneurysm} {Repair} {Decision} {Aid} and economic
  evaluation,'' {\em Health Technology Assessment}, vol.~19, pp.~1--154, v--vi,
  Apr. 2015.

\bibitem{bellman:1957}
R.~Bellman, {\em Dynamic Programming}.
\newblock Princeton, NJ, USA: Princeton University Press, 1957.

\bibitem{magni_deciding_2000}
P.~Magni, S.~Quaglini, M.~Marchetti, and G.~Barosi, ``Deciding when to
  intervene: a {Markov} decision process approach,'' {\em International Journal
  of Medical Informatics}, vol.~60, no.~3, pp.~237--253, 2000.

\bibitem{Scott2016}
S.~W.~M. Scott, A.~J. Batchelder, D.~Kirkbride, A.~R. Naylor, and J.~P.
  Thompson, ``{Late Survival in Nonoperated Patients with Infrarenal Abdominal
  Aortic Aneurysm.},'' {\em European Journal of Vascular and Endovascular
  Surgery}, vol.~52, pp.~444--449, oct 2016.

\end{thebibliography}

\appendices

\section{Regarding the rupture risk for large \\ ($> 55$ mm) abdominal aortic aneurysms}
\label{app:big_AAA}

As mentioned in the results section of this paper (i.e.,
Section~\ref{sec:results}), the abrupt change in the policy at 55 mm is somewhat
unexpected. We mentioned that such a discontinuity likely corresponds to a bias
in the data. In particular, that either the rupture risk for large aneurysms is
estimated as disproportionally high, or that the rupture risk for small aneurysms is
estimated as disproportionally low.

After publication of this manuscript, it has been proposed in \cite{Scott2016},
that the rupture risks presented in the literature for large aneurysms are
overestimates. In particular, \cite{Scott2016} indicates that rupture risks for
aneurysms between 55 and 69 mm are overestimated. With respect to our results,
this would lead to a bias in the policy, such that larger aneurysms are
disproportionally avoided (i.e., operated). This would manifest itself as an abrupt
change in the policy for aneurysms of size 55 mm and larger -- exactly as is
apparent in Figure~\ref{fig:policies}. Thus, we believe that our results also point
to possible previous overestimation of rupture risks for large aneurysms.

Additionally, we note that the sensitivity analysis presented in this paper
only demonstrates the stability of the policy with respect to variance in the
parameters and would therefore not be able to detect the systematic bias
proposed in \cite{Scott2016}.

\end{document}